  \documentstyle[12pt]{article}
  
\catcode`@=11
\def\secteqno{\@addtoreset{equation}{section}%
\def\theequation{\thesection.\arabic{equation}}}
\catcode`@=12
\secteqno
\newcommand{\be}{\begin{equation}}
\newcommand{\ee}{\end{equation}}
\newcommand{\bea}{\begin{eqnarray}}
\newcommand{\eea}{\end{eqnarray}}
\newcommand{\bref}[1]{(\ref{#1})}
\newcommand{\nn}{\nonumber}

\def\ads{{AdS$_5\times$S$^5~$}} 
\def\CP{{\frac{1+\Gamma_{11}}{2}}}

\def\RS{{R'}}
\begin{document}
\thispagestyle{empty}
\vfill
\hfill February 27 , 2002\par
\hfill KEK-TH-805\par
\hfill TOHO-CP-0270\par
\vskip 20mm
\begin{center}
{\Large\bf From Super-\ads Algebra}\\
{\Large \bf to Super-pp-wave Algebra}\par
\vskip 6mm
\medskip

\vskip 10mm
{\large Machiko\ Hatsuda,~Kiyoshi\ Kamimura$^\dagger$~and~Makoto\ Sakaguchi 
}\par
\medskip
{\it 
Theory Division,\ High Energy Accelerator Research Organization (KEK),\\
\ Tsukuba,\ Ibaraki,\ 305-0801, Japan} \\
{\it 
$~^\dagger$ 
 Department of Physics, Toho University, Funabashi, 274-8510 Japan}\\
\medskip
{\small E-mails:\ mhatsuda@post.kek.jp, kamimura@ph.sci.toho-u.ac.jp,
Makoto.Sakaguchi@kek.jp} 
\medskip
\end{center}

\vskip 10mm
\begin{abstract}
The isometry algebras of the maximally supersymmetric solutions of IIB
supergravity are derived 
by the In$\ddot{\rm o}$n$\ddot{\rm u}$-Wigner 
 contractions of the super-\ads algebra. 
The super-\ads algebra allows introducing two contraction parameters;
the one for the Penrose limit to the maximally supersymmetric
pp-wave algebra
and the \ads radius for the flat limit.
The fact that the Jacobi identity of three supercharges holds 
irrespectively 
of these parameters reflects the fact that the number of supersymmetry is not affected
 under both contractions.
\end{abstract} 
\par\par
\newpage
\setcounter{page}{1}
\parskip=7pt
\section{ Introduction}\par
\indent

Recently the supersymmetric pp-wave backgrounds, as well as the anti-de Sitter (AdS) 
backgrounds, have been widely studied as supergravity vacua.
A maximally supersymmetric pp-wave solution for the eleven-dimensional 
supergravity theory was found in \cite{KG1,CHM,KG2,joseM},
and one for the type IIB supergravity theory was found in \cite{joseIIB}.
Relations between the pp-wave background and the AdS background
 are crucial for the point of view of the 
AdS/CFT correspondence \cite{Me1,Malda,MeTsy2}
where the pp-wave background is recognized as an approximation of the AdS background.
The plane wave space is obtained as the limiting space from arbitrary space 
shown first by Penrose
\cite{Penrose},
and this idea is extended to supergravity and superstring theories \cite{Guven,joseP,joseST}\footnote{This limiting procedure had appeared in the context of the 
Wess-Zumino-Witten model in \cite{sfe1}.}.

The maximally supersymmetric solution of the IIB supergravity consist of the 
pp-wave metric and the self-dual null homogeneous 5 form flux \cite{joseIIB}
\bea
ds^2&=&2dx^-dx^+-4\lambda^2\sum_{\hat i=1}^{8}(x^{\hat i})^2(dx^-)^2+
\sum_{\hat i=1}^{8}(dx^{\hat i})^2
\nn\\
F_5&=&\lambda~dx^-\wedge(dx^1\wedge dx^2\wedge dx^3\wedge dx^4+
                         dx^5\wedge dx^6\wedge dx^7\wedge dx^8)
\label{ppmetric}
\eea
with a constant dilaton. 
It has 32 Killing spinors preserving the solution. 
Recently it was discussed 
in detail that the solution \bref{ppmetric} is obtained as
a Penrose limit of the \ads  background \cite{joseP,joseST}. 
Since the pp-wave solution \bref{ppmetric} is obtained as
a limiting case  of the \ads solution by the Penrose limit \cite{Penrose}, 
the isometry algebra of the former should be also obtained from the latter 
by some limiting procedure. 
The Penrose limit is usually discussed as a mapping between metrics
in terms of local coordinates.
By using canonical symmetry generators written in terms of local coordinates, 
this mapping directly tells the relation of 
generators, thus of algebras.

In this paper we examine 
an In$\ddot{\rm o}$n$\ddot{\rm u}$-Wigner (IW) contraction
\cite{IW} of algebra which is independent of choice of local coordinates. 
Then the Penrose limit maps not only the \ads metric into pp-wave metric 
but also the super-\ads algebra into  the maximally supersymmetric 
pp-wave algebra\footnote{We refer the isometry algebra of the pp-wave 
metric as the pp-wave algebra.}
as an IW contraction.
The IW contraction from the AdS algebra to 
the pp-wave algebra for the bosonic case is discussed in the section 2,
and the one for the supersymmetric case is in the section 3.\medskip

\section{ From AdS algebra to pp-wave algebra}\par
\indent

The AdS$_d\times$S$^{D-d}$ algebra is given, in terms of dimensionless 
momenta $P$'s and rotations $M$'s, as
\bea
\begin{array}{lcl}
\left[P_a,P_b\right]=M_{ab}&,&\left[P_{a'},P_{b'}\right]=-M_{a'b'}\\
\left[P_a,M_{bc}\right]=\eta_{ab}P_c-\eta_{ac}P_b&,&
\left[P_{a'},M_{b'c'}\right]=\eta_{a'b'}P_{c'}-\eta_{a'c'}P_{b'} \\
\left[M_{ab},M_{cd}\right]=\eta_{bc}M_{ad}+ 3{~\rm terms}&,&
\left[M_{a'b'},M_{c'd'}\right]=\eta_{b'c'}M_{a'd'}+ 3{~\rm terms}\\
\end{array}
\label{bosonads}\eea
where $a=0,1,...,d-1~$ and $a'=d,...,D-1~$ stand for vector indices  of  AdS$_d$ 
and S$^{D-d}$ respectively.
The symmetry group is isomorphic to SO$(d-1,2)\times$SO$(D-d+1)$
which has a flat limit 
to ISO$(d-1,1)\times$ ISO$(D-d)$ 
by the following IW contraction \cite{IW}. 
\begin{enumerate}
\item Rescale the translation generators $P$'s using 
the radii of the AdS$_d$ and S$^{D-d}$ , $R$ and $\RS$ respectively, as
\bea
P_a\to RP_a,~~~~~P_{a'}\to \RS P_{a'}.
\label{scalingR}
\eea
\item Then take $R\to \infty$ and $\RS\to \infty$.  
\end{enumerate}  
In the limit, $P$'s become the linear momenta and $M$'s are the Lorentz
generators. Taking large limit of the radii, $R$ and $R'$, 
corresponds to reducing an AdS$\times$S space into a flat space.
\vskip 3mm

Besides the flat limit, as any other metrics
the AdS$\times$S metric allows a limit giving a plane wave metric 
 ( Penrose limit ) \cite{Penrose}. 
The Penrose limit 
can be understood as an IW contraction of the AdS$\times$S algebra
into the pp-wave algebra. 
\begin{enumerate}
\item Define the light cone components of the momenta $P$ 
and {\it boost} generators $P^*$ as
\bea
P_{\pm}&\equiv&\frac{1}{\sqrt{2}}( P_{D-1}\pm P_0),~~~
P_{\hat{i}}~=~\left\{\begin{array}{l}P_i\\P_{i'}\end{array}\right\},~~~
P_{\hat{i}}^*=\left\{\begin{array}{l}P_i^*\equiv M_{0i}\\P_{i'}^*
\equiv M_{(D-1)i'}\end{array}\right\}.
\label{lc}
\eea
where $\hat{i}=1,.....,D-2,~i=1,...,d-1,~i'=d,...,D-2.$
\item
Suppose the plane wave propagates with respect to $x_+$ {\it time}.
The transverse translation and {\it boost} generators
are rescaled with
the dimensionless parameter $\Omega$ 
as
\bea
P_+\to \frac{1}{\Omega^2}P_+~~,
P_{\hat{i}}\to \frac{1}{\Omega}P_{\hat{i}}~~,~~
P_{\hat{i}}^*\to \frac{1}{\Omega}P_{\hat{i}}^*.
\label{penomg}
\eea
\item Then take $\Omega\to 0$ limit.  
\end{enumerate}  
In the limit, $P_+$ becomes a center, and it can be treated as a constant.

To see them explicitly the AdS$\times$S algebra \bref{bosonads} 
is rescaled following to \bref{scalingR} and \bref{penomg}
\bea
\begin{array}{lcl}
\left[P_+,P_i\right]=\frac{\Omega^2}{\sqrt{2}R^2}P_i^*&,&\left[P_-,P_i\right]=-\frac{1}{\sqrt{2}R^{2}}P_i^*\\
\left[P_+,P_{i'}\right]=-\frac{\Omega^2}{\sqrt{2}\RS^{2}}P_{i'}^*&,&
\left[P_-,P_{i'}\right]=-\frac{1}{\sqrt{2}\RS^{2}}P_{i'}^*\\ 
\left[P_+,P_{i}^*\right]=-\frac{\Omega^2}{\sqrt{2}}P_i&,&
\left[P_-,P_{i}^*\right]=\frac{1}{\sqrt{2}}P_i\\
\left[P_+,P_{i'}^*\right]=\frac{\Omega^2}{\sqrt{2}}P_{i'}&,&
\left[P_-,P_{i'}^*\right]=\frac{1}{\sqrt{2}}P_{i'}\\
&&\\
\left[P_i,P_{j}\right]=\frac{\Omega^2}{R^{2}}M_{ij}&,&
\left[P_{i'},P_{j'}\right]=-\frac{\Omega^2}{\RS^{2}}M_{i'j'}\\ 
\left[P_i^*,P_{j}^*\right]={\Omega^2}M_{ij}&,&
\left[P_{i'}^*,P_{j'}^*\right]=-{\Omega^2}M_{i'j'}\\
\left[P_i,P_{j}^*\right]=-\frac{1}{\sqrt{2}}\eta_{ij}(P_+-\Omega^2 P_-)&,&
\left[P_{i'},P_{j'}^*\right]=-\frac{1}{\sqrt{2}}\eta_{i'j'}(P_++\Omega^2 P_-)\\
&&\\
\left[P_{\hat{i}},M_{\hat{j}\hat{k}}\right]=\eta_{\hat{i}[\hat{j}}P_{\hat{k}]}&,&
\left[P_{\hat{i}}^*,M_{\hat{j}\hat{k}}\right]=
\eta_{\hat{i}[\hat{j}}P_{\hat{k}]}^*\\
\left[M_{\hat{i}\hat{j}},M_{\hat{k}\hat{l}}\right]
=\eta_{\hat{k}\hat{j}}M_{\hat{i}\hat{l}}+3 {\rm terms}&,&
{\rm others}=0.
\end{array}\label{bosonRomg}
\eea
By taking $\Omega\to 0$ limit,
\bref{bosonRomg} becomes the plane wave algebra, 
\bea
&&\left[P_-,P_{\hat{i}}^*\right]=\frac{1}{\sqrt{2}}P_{\hat{i}}~~,~~
\left[P_-,P_{\hat{i}}\right]=-4\sqrt{2}{\lambda_{\hat{i}}^{2}}P_{\hat{i}}^*~~,~~
\left[P_{\hat{i}},P_{\hat{j}}^*\right]=-\frac{1}{\sqrt{2}}\eta_{\hat{i}\hat{j}}P_+\nn\\
&&\left[P_{\hat{i}},M_{\hat{j}\hat{k}}\right]=\eta_{\hat{i}[\hat{j}}
P_{\hat{k}]}~~,~~
\left[P_{\hat{i}}^*,M_{\hat{j}\hat{k}}\right]=
\eta_{\hat{i}[\hat{j}}P_{\hat{k}]}^*~\label{bosonpp}\\
&&\left[M_{\hat{i}\hat{j}},M_{\hat{k}\hat{l}}\right]=
\eta_{\hat{k}\hat{j}}M_{\hat{i}\hat{l}}+3 {\rm terms}
~~,~~{\rm others}=0\nn
\eea
where
\bea
\lambda_{\hat{i}}=\left\{\begin{array}{l}\lambda_i=
\frac{1}{2\sqrt{2}R}\\
\lambda_{i'}=
\frac{1}{2\sqrt{2}\RS}\end{array}\right.~~.
\eea
This is the symmetry algebra of the pp-wave metric 
\cite{Krm}.
$P_+$ is a central element in \bref{bosonpp}
allowing arbitrary rescaling
\footnote{In \cite{joseM,joseIIB} $P_+$ is rescaled as $\lambda^2 P_+$,
and $P$'s and $P^*$'s  are 
$e^*$'s and $e$'s  respectively.
Or $e^*$'s can be recognized as $\lambda^{-2}e^*$ corresponding to $P^*$,
but this prohibits the flat limit $\lambda\to 0$ in their superalgebra. } with $\lambda$.

Furthermore the flat limit is taken by
$R,\RS\to \infty$ in \bref{bosonRomg} or 
$\lambda_{\hat{i}}\to 0$  in \bref{bosonpp}. 
Although this flat limit algebra is different from 
the direct flat limit from the AdS algebra, 
the $D$-dimensional Poincare group ISO($D-1,1$)
can be recovered 
 by supplying spontaneously broken generators consistently.

\par\vskip 6mm


\section{ From Super-\ads algebra to Super-pp wave algebra}\par
\indent

We extend the previous analysis
to the supersymmetric case 
and show the maximally supersymmetric IIB pp-wave algebra \cite{joseIIB}
is obtained by a contraction of the super-\ads algebra. In this case
the Jacobi identities of the superalgebra give a restriction 
on the scale parameters $R$ and $\RS$ in \bref{bosonads}, as $R=\RS$.
We start with the super-\ads algebra following to 
the notation of Metsaev and Tseytlin \cite{MeTsy}. 
In 10 dimensional IIB theory the supersymmetry generators $Q_A$
are chiral Majorana spinors and are SL(2,R) doublet
$( A=1,2)$. The gamma matrices $\Gamma_{\hat a},~({\hat a}=0,1,..,9)$ 
in 10 dimensions are composed using
those of AdS$_5$, $\gamma_a,~({a}=0,1,..,4)$, and of S$^5$, $\gamma_a',~
({a'}=5,..,9)$, as
\bea
&&\Gamma_a=\gamma_a\otimes {\bf 1}'\otimes \sigma_1~~,~~
\Gamma_{a'}={\bf 1}\otimes \gamma_{a'}\otimes \sigma_2~~,~~
\eea
and satisfy
\bea
\begin{array}{lcl}
\{\Gamma_{\hat{a}}\Gamma_{\hat{b}}\}=2\eta_{\hat{a}\hat{b}}
&,&\eta_{\hat{a}\hat{b}}={(-++++,+++++)}\nn\\
\{\gamma_{a},\gamma_{b}\}=2\eta_{ab}&,&\eta_{ab}=(-++++)\\
\{\gamma_{a'},\gamma_{b'}\}=2\eta_{a'b'}&,&
\eta_{a'b'}=(+++++).
\end{array}\label{eta}
\eea
The supercharges $Q_A$'s are satisfying 
\bea
Q_A=Q_A~\CP
~~,~~
\Gamma_{11}=\Gamma_{0123456789}={\bf 1}\otimes {\bf 1}'\otimes \sigma_3~~~.
\eea
The charge conjugation matrix ${\cal C}$ in 10 dimensions is taken as
\bea
{\cal C}=C\otimes C'\otimes i\sigma_2
\eea
where $C$ and $C'$ are respectively 
the charge conjugations for AdS$_5$ and S$^5$ spinors.

The bosonic part of supersymmetric \ads algebra is \bref{bosonads} 
with $d=5, D=10$. In addition to it the odd generators $Q_A$ satisfy
\bea
\left[Q_A,P_a\right]=\frac{i}{2}Q_{B}\gamma_{a}\epsilon_{BA}&,&
\left[Q_{A},P_{a'}\right]=-\frac{1}{2}Q_B\gamma_{a'}\epsilon_{BA}\nn\\
\left[Q_A,M_{ab}\right]=-\frac{1}{2}Q_{A}\gamma_{ab}&,&
\left[Q_{A},M_{a'b'}\right]=-\frac{1}{2}Q_A\gamma_{a'b'}\nn
\eea
\bea
\left\{Q_{\alpha\alpha'A},Q_{\beta\beta'B}\right\}&=&
\delta_{AB}\left[
-2i{C'}_{\alpha'\beta'}(C\gamma^a)_{\alpha\beta}P_a
+2C_{\alpha\beta}(C'\gamma^{a'})_{\alpha'\beta'}P_{a'}
\right]
\label{QQ}\\
&+&
\epsilon_{AB}\left[
{C'}_{\alpha'\beta'}(C\gamma^{ab})_{\alpha\beta}M_{ab}
-C_{\alpha\beta}(C'\gamma^{a'b'})_{\alpha'\beta'}M_{a'b'}
\right]
\nn~~
\eea
where we omit to write the chiral projection operators.

For the present purpose we rewrite this algebra 
in terms of 10-dimensional covariant gamma matrices, 
$\Gamma$'s, rather than $\gamma$'s.
It is convenient to introduce following matrices  
\bea
&&\Gamma_{01234}={\bf 1}\otimes {\bf 1}'\otimes -i\sigma_1\equiv\Gamma_0  
I~~,~~~~~~~~~
\Gamma_{56789}={\bf 1}\otimes {\bf 1}'\otimes \sigma_2\equiv\Gamma_9 
J~~~.\label{gammabar}
\eea
The AdS superalgebra \bref{QQ} are rewritten as 
\bea
\left[Q_A,P_{\hat{a}}\right]&=&-\frac{1}{2}Q_{B}\Gamma_0I\Gamma_{\hat{a}}\epsilon_{BA}\nn\\
\left[Q_A,M_{\hat{a}\hat{b}}\right]&=&-\frac{1}{2}Q_{A}\Gamma_{\hat{a}\hat{b}}
\label{QQG}\\
\left\{Q_{\hat{\alpha}A},Q_{\hat{\beta}B}\right\}&=&
-2i\delta_{AB}{\cal C}\Gamma^{\hat{a}}P_{\hat{a}}
+i\epsilon_{AB}
\left[{\cal C}\Gamma^{{a}{b}}\Gamma_0I M_{{a}{b}}+
{\cal C}\Gamma^{{a'}{b'}}\Gamma_9 J
M_{{a'}{b'}}\right]
\nn~~
\eea
using with the useful relation in the first line
\bea
&&
\CP(\Gamma_0 I)=\CP(\Gamma_9 J)~~.\label{0i9j}
\eea
\vskip 3mm

Associating with the rescaling of $P$'s by $R$ as in \bref{scalingR}
the dimensionless supercharge $Q_A$ is rescaled as
\bea
P_a\to RP_a,~P_{a'}\to R P_{a'}
\label{scalingR2}~~,~~
Q_A\to \sqrt{R}~Q_A.
\label{scalingQR}
\eea
Corresponding to the rescaling of bosonic generators \bref{penomg}
the components of the supercharges are rescaled with proper weights.
The $Q_A$ is decomposed as 
\bea
Q_A&=&Q_{+A}+Q_{-A}~~,~~~~~Q_{\pm A}\Gamma_\mp~\equiv~0~~,~~
\Gamma_\pm\equiv\frac{1}{\sqrt{2}}(\Gamma_9\pm\Gamma_0)~~.
\eea
The supercharges 
are rescaled  as
\bea
Q_+\to \frac{1}{\Omega}~Q_+~~,~~Q_-\to Q_-
\label{penQomg}
\eea
in  the superalgebra of the AdS$^5\times$S$^5$ 
to obtain the well defined Penrose limit.

The super-\ads algebra \bref{QQG}  after the rescaling of the 
$P$'s, \bref{scalingR} with $R=\RS$ and \bref{penomg}, and those of $Q$'s
, \bref{scalingQR} and \bref{penQomg},
becomes
\bea
\begin{array}{lcl}
\left[Q_{+,A},P_{-}\right]=-\frac{1}{\sqrt{2}R}Q_{+,B}I\epsilon_{BA}
&,&\left[Q_{-,A},P_{+}\right]=\frac{\Omega^2}{\sqrt{2}R}Q_{-,B}I\epsilon_{BA}\\
\left[Q_{+,A},P_{\hat{i}}\right]=\frac{\Omega^2}{2\sqrt{2}R}
Q_{-,B}{\Gamma}_{-}I\Gamma_{\hat{i}}\epsilon_{BA}&,&
\left[ Q_{-,A},P_{\hat{i}} \right]=-\frac{1}{2\sqrt{2}R}Q_{+,B}\Gamma_+I\Gamma_{\hat{i}}\epsilon_{BA}
\\
\left[Q_{+,A},P_{i}^*\right]=\frac{\Omega^2}{2\sqrt{2}} Q_{-,A}\Gamma_-\Gamma_{i}&,&\left[Q_{-,A},P_{\hat{i}}^*\right]= -\frac{1}{2\sqrt{2}}
Q_{+,A}\Gamma_+\Gamma_{\hat{i}}\\
\left[Q_{+,A},P_{i'}^*\right]=-\frac{\Omega^2}{2\sqrt{2}}
 Q_{-,A}\Gamma_-\Gamma_{\hat{i}}&,&
\left[Q_{\pm,A},M_{\hat{i}\hat{j}}\right]=-\frac{1}{2}Q_{\pm,A}\Gamma_{\hat{i}\hat{j}}\\
{\rm others}=0&&
\end{array}\label{susyome}
\eea
\bea
\left\{Q_{+,A},Q_{+,B}\right\}&=&
-2i\delta_{AB}{\cal C}\Gamma^+P_+
-\frac{i\Omega^2}{\sqrt{2}R} \epsilon_{AB}
\left[{\cal C}\Gamma^+\Gamma^{{i}{j}}I M_{{i}{j}}
-{\cal C}\Gamma^+\Gamma^{{i'}{j'}}J M_{{i'}{j'}}\right]\label{susyomeQQ}
\label{susyomgqq}
\\
\left\{Q_{+,A},Q_{-,B}\right\}&=&
-2i\delta_{AB}{\cal C}\Gamma^{\hat{i}}\frac{\Gamma_-\Gamma_+
}{2}
P_{\hat{i}}
-\frac{2i}{R}\epsilon_{AB}\left[ {\cal C}\frac{\Gamma_-\Gamma_+
}{2}\Gamma^{i} I P_{i}^*
+{\cal C}\frac{\Gamma_-\Gamma_+
}{2}\Gamma^{i'} J P_{i'}^*\right]\nn
\\
\left\{Q_{-,A},Q_{-,B}\right\}&=&
-2i\delta_{AB}{\cal C}\Gamma^{-}P_{-}
+\frac{i}{\sqrt{2}R}\epsilon_{AB}
\left[ {\cal C}{\Gamma}^{-}\Gamma^{ij}I M_{ij}
+{\cal C}{\Gamma}^{-}\Gamma^{i'j'}J M_{i'j'}
\right]~~.\nn
\eea
Here we have used relations following from \bref{0i9j}, 
\bea
\Gamma_\pm I~\CP=
\mp~\Gamma_\pm J~\CP~~.~
\eea
It is important that negative power terms of $\Omega$ disappear
due to the presence of the chiral and the light cone projections.
Therefore we can take the Penrose limit $\Omega\to 0$ of the algebra to obtain
\bea
\begin{array}{lcl}
\left[Q_{A},P_{-}\right]=-{\lambda} Q_{B}(I+J)\epsilon_{BA}
&,&\left[Q_{A},P_{i}\right]={\lambda} Q_{B}{\Gamma}_{+}\Gamma_{i}I\epsilon_{BA}\\
\left[Q_{A},P_{\hat{i}}^*\right]=-\frac{1}{2\sqrt{2}}Q_{A}\Gamma_+\Gamma_{\hat{i}}&,&
\left[Q_{A},P_{i'}\right]={\lambda} Q_{B}{\Gamma}_{+}\Gamma_{i'}J\epsilon_{BA}
\label{QPpp}\\
\left[Q_{A},M_{\hat{i}\hat{j}}\right]=-\frac{1}{2}Q_{A}\Gamma_{\hat{i}\hat{j}}&&
\end{array}
\eea
\bea
\left\{Q_{A},Q_{B}\right\}&=&
-2i\delta_{AB}\left[
{\cal C}\Gamma^+P_++{\cal C}\Gamma^{-}P_{-}+{\cal C}\Gamma^{\hat{i}}P_{\hat{i}}
\right]\nn\\
&&+4\sqrt{2}i\lambda \epsilon_{AB}
\left[{\cal C}I\Gamma^{i}P_{i}^*+{\cal C}J\Gamma^{i'}P_{i'}^*
\right]\label{QQpp}\\
&&+2i\lambda \epsilon_{AB}
\left[{\cal C}\Gamma^{-}I\Gamma^{ij}M_{ij}+
{\cal C}\Gamma^{-}J\Gamma^{i'j'}M_{i'j'}
\right]\nn
\eea
where 
$\lambda={1}/{(2\sqrt{2}R)}$. 
This is the maximally supersymmetric pp-wave algebra obtained in 
\cite{joseIIB}.

Furthermore the flat limit is taken by
$R\to \infty$ in \bref{susyome} and \bref{susyomeQQ}
 or $\lambda\to 0$  in  \bref{QQpp}.

\section{ Summary and Discussions}\par
\indent

In this paper we have derived the super-pp-wave algebra
taking with 
an IW contraction of the super-\ads algebra.
It is stressed that the Jacobi identities of this algebra 
\bref{bosonRomg}, \bref{susyome} and \bref{susyomeQQ} hold for any value of  
$\Omega$ and $\lambda$, and the algebra has well defined
even in their zero limits.
It explains naturally why the pp-wave and the flat supersymmetry 
are maximally supersymmetric as the super-\ads.
The number of bosonic generators in \ads and pp-wave algebra
are same since their algebras are connected by the contraction.

The relation between the super-AdS background and the super-pp-wave  background
at algebraic level is practical to construct mechanical actions 
for branes in the super-pp-wave background.
Any form field in the latter can be obtained from the former.
This approach manifests symmetries of super-pp-wave systems
whose dynamics and spectrum will be analyzed elegantly. 
\par

\medskip


\end{document}